\documentclass[aoas,nameyear,seceqn,dvips]{arximspdf}
\usepackage{graphicx}

\doi{10.1214/10-AOAS396}
\volume{5}
\issue{2A}
\pubyear{2011}
\firstpage{1102}
\lastpage{1125}

\makeatletter

\newproclaim{remarks}{Remarks}

\def\eqref#1{(\ref{#1})}

\makeatother

\begin{document}
\begin{frontmatter}

\title{The generalized shrinkage estimator for the analysis of
functional connectivity of brain~signals\thanksref{TT}}
\runtitle{The generalized shrinkage estimator}
\thankstext{TT}{Supported in part by the National Science Foundation
(Division of Mathematical Sciences).}
\begin{aug}
\author[A]{\fnms{Mark} \snm{Fiecas}\ead[label=e1]{mfiecas@stat.brown.edu}}
\and
\author[A]{\fnms{Hernando} \snm{Ombao}\corref{}\ead[label=e2]{ombao@stat.brown.edu}}
\runauthor{M. Fiecas and H. Ombao}
\affiliation{Brown University}
\address[A]{Center for Statistical Sciences\\
Brown University\\
121 S. Main Street, 7th Floor\\
Providence, Rhode Island 02912\\
USA\\
\printead{e1}\\
\phantom{E-mail: }\printead*{e2}}
%Main Street, 7th Floor \\ Providence, Rhode Island 02912 \\ USA\\
%}
\end{aug}

% HISTORY:
\received{\smonth{3} \syear{2010}}
\revised{\smonth{8} \syear{2010}}

% ABSTRACT
%
\begin{abstract}
We develop a new statistical method for estimating functional
connectivity between neurophysiological signals represented
by a multivariate time series. We use partial coherence as the measure
of functional connectivity. Partial coherence
identifies the frequency bands that drive the direct linear association
between any pair of channels. To estimate partial
coherence, one would first need an estimate of the spectral density
matrix of the multivariate time series. Parametric
estimators of the spectral density matrix provide good frequency
resolution but could be sensitive when the parametric
model is misspecified. Smoothing-based nonparametric estimators are
robust to model misspecification and are consistent
but may have poor frequency resolution.
In this work, we develop the generalized shrinkage estimator, which is
a weighted average of a parametric estimator and a
nonparametric estimator. The optimal weights are frequency-specific and
derived under the quadratic risk criterion so that
the estimator, either the parametric estimator or the nonparametric
estimator, that performs better at a particular frequency
receives heavier weight. We validate the proposed estimator in a
simulation study and apply it on electroencephalogram
recordings from a visual-motor experiment.
\end{abstract}

\begin{keyword}
\kwd{Multivariate time series}
\kwd{periodogram matrix}
\kwd{shrinkage}
\kwd{spectral density matrix}
\kwd{vector autoregressive model}.
\end{keyword}

\end{frontmatter}

%s1 ###
\section{Introduction}
The goal of this paper is to estimate dependence between multi-channel
electroencephalogram (EEG) signals. In the time domain, partial
cross-correlation is used to measure the strength of direct linear
dependence between a pair of channels. In the frequency domain, partial
coherence is utilized to identify the frequency bands that drive the
direct linear association. To estimate partial coherence, one first
needs to estimate the spectral density matrix, which is done via a
parametric method (by fitting a~parametric model
to the EEGs) or by some nonparametric procedure such as
kernel-smoothing. Both of these procedures have their strengths and
weaknesses. In this paper we develop a generalized shrinkage procedure
which is a weighted average of the parametric and nonparametric
estimates. The frequency-specific weights are derived data-adaptively
so that the estimator (parametric versus nonparametric) that performs
better at a particular frequency receives heavier weight.

To obtain a parametric estimate of the spectral density matrix, one can
fit a model, such as the vector autoregressive model (VAR), which has been
applied extensively in the analysis of a variety of brain signals
[e.g., \citet{Goebel03}; \citet{Eichler05}; \citet
{Schlogl06}; \citet{Thompson09}]. It is simple
and can be easily applied for assessing Granger-causality and the
direction of information between the signals [\citet{Kaminski91};
\citet{Kaminski01}]. Moreover, when the lag order is sufficiently
large, the
VAR estimates are known to be well localized in frequency.
Nonparametric estimators are derived from periodogram matrices which
are the data analog of the spectral density matrix. One nonparametric
estimator is obtained by smoothing the periodograms across frequencies.
The performance of these estimators is a function of the smoothing
parameter and so these estimators may not always give well-localized
estimates. However, they are asymptotically mean squared consistent and
robust to model specification. In this work we develop the generalized
shrinkage estimator which is a weighted average of the parametric
estimator and the nonparametric estimator and thus provides a good
compromise between these two estimators.

The remainder of the paper is organized as follows. In Section \ref
{Background} we discuss partial coherence and the past works on
shrinkage estimation. Section \ref{Methods} lays out the framework for
the generalized shrinkage estimator. In Section \ref{Simulations} we
show the performance of the generalized shrinkage estimator relative to
the VAR estimator, smoothed periodogram and the multitaper on a
simulated data set. In Section \ref{EEG} we use the generalized
shrinkage estimator to analyze functional connectivity on an EEG data
set. And finally, in Section \ref{Discussion}, we summarize the
conclusions of this research and briefly discuss properties of the
generalized shrinkage estimator and future directions.

%s2 ###
\section{Background}\label{Background}
Coherence, the frequency domain analog of cross-corre\-lation, is a
temporally invariant frequency-specific measure of linear asso\-ciation
between signals. Consider a trivariate time series $(X(t), Y(t),
Z(t))^\top$. Denote $X_{\omega}(t), Y_{\omega}(t)$ and $Z_{\omega
}(t)$ to be the bandpass-filtered signals so that each of their spectra
is concentrated on some narrow frequency band around~$\omega$. \citet
{Ombao08} showed that coherence has an appealing interpretation of
being the squared absolute cross-correlation between a pair of bandpass
filtered signals, so that in this case, the coherence between $X(t)$
and $Y(t)$ at frequency $\omega$ can be interpreted as the squared
cross-correlation between $X_{\omega}(t)$ and $Y_{\omega}(t)$.
However, if the linear association between $X(t)$ and $Y(t)$ at
frequency $\omega$ is confounded by $Z(t)$, conclusions based on the
associations between $X(t)$ and $Y(t)$ may be misleading and
misinterpreted because they may be related only indirectly through~$Z(t)$.
In other words, coherence does not model \textit{direct}
linear association. We can obtain a \textit{direct} measure of
frequency-specific linear association using {\it partial coherence},
which is interpreted as the squared cross-correlation between
$X_{\omega}(t)$ and~$Y_{\omega}(t)$ at frequency $\omega$ after
removing the temporally invariant linear effects of~$Z(t)$. To estimate
partial coherence, we use a characterization that expresses partial
coherence as a function of the inverse of the spectral density matrix
[\citet{Dahlhaus00}]. In fact, this characterization was used in
\citet
{Eichler03} for neural spike trains, \citet{Salvador05} for fMRI, and
\citet{Medkour09} for EEG.

%s2.1 ###
\subsection{Characterization of partial coherence}\label{obtainPCoh}
$\!\!$Let ${\mathbf{X}}(t) \!=\! (X_1(t), \ldots, X_P(t))^\top$ be a
$P$-dimensional weakly stationary zero-mean real-valued discrete time
series with spectral density matrix ${\mathbf{f}}(\omega)$. The
diagonal elements of ${\mathbf{f}}(\omega)$, denoted $f_{pp}(\omega
)$, $p=1, \ldots, P$, are the autospectra of the $P$ channels and each
of the off-diagonal elements, denoted $f_{pq}(\omega), p \neq q$, is
the cross-spectrum between channels $X_p(t)$ and $X_q(t)$. Define the
matrix ${\mathbf{g}}(\omega) = {\mathbf{f}}^{-1}(\omega)$ whose
$(p,q)$th element is denoted as $g_{pq}(\omega)$. Let ${\mathbf
{h}}(\omega)$ be a diagonal matrix whose elements are
$g_{pp}^{-1/2}(\omega)$. Define the matrix $\Gamma(\omega)$ to be
%
%e2.1 ###
\begin{eqnarray} \label{Eq:Gamma}
\Gamma(\omega) = - {\mathbf{h}}(\omega) \mathbf{g}(\omega)
{\mathbf{h}}(\omega).
\end{eqnarray}
Then the partial coherence $\rho_{pq}(\omega)$ between the $p$ and
$q$th channels at frequency $\omega$ is the square of the modulus of
the $(p,q)$th element of $\Gamma(\omega)$, that is, $\rho
_{pq}(\omega) = |\Gamma_{pq}(\omega)|^2$.
To estimate partial coherence, we must first estimate the spectral
density matrix.

%s2.2 ###
\subsection{Related work on shrinkage estimation}
\citet{Ledoit04} proposed shrinkage estimation of the
variance--covariance matrix that combines a ``classical estimator'' (the
sample variance--covariance matrix) with a ``highly-structured target.''
Recently, the idea of a convex combination of a ``classical estimator''
with a target has been extended for estimating the spectral density
matrix of a multivariate time series, which is the frequency-domain
analog of the variance--covariance matrix. \citet{Bohm09}
developed the
shrinkage estimator for the spectral density matrix which shrinks the
nonparametric estimator to the scaled identity matrix.
When shrinking the nonparametric estimator toward the scaled identity
matrix the resulting estimate is well conditioned.
However, the off-diagonals of the estimator is shrunk to 0 and thus
potentially biases the estimates of linear association
toward the null.
To overcome this problem, one may shrink the nonparametric estimator
toward a more general shrinkage target.
For factor models in economic time series, \citet{Bohm08}
proposed to
shrink the nonparametric estimator toward a structured model, namely,
the one-factor model. Here, we extend these works by giving the
shrinkage weight for any arbitrary shrinkage target.

%s3 ###
\section{The generalized shrinkage estimator}\label{Methods}
Let $\mathbf{X}_n(t)$, $n = 1, 2, \ldots, N$ and $t = 1, 2, \ldots,
T$, be the $n$th trial of a $P$-dimensional weakly stationary zero-mean
real-valued discrete time series with auto-covariance
matrix $\Sigma(h)$, each of whose elements is absolutely summable. We
shall assume that the trials are independent realizations from a common
underlying process whose spectral properties (in particular, partial
coherence) we wish to investigate. To achieve the
goal of estimating partial coherence, we shall first estimate the
$P\times P$ spectral density matrix defined by
$\mathbf{f}(\omega)= \frac{1}{2\pi}\sum_{h \in\mathbb{Z}} \Sigma
(h) \exp
(-i\omega h)$.
Denote the parametric estimator of $\mathbf{f}(\omega)$ to be
$\widetilde{\mathbf{V}}(\omega)$ and the
nonnparametric estimator to be $\widetilde{\mathbf{f}}(\omega)$.
The generalized shrinkage estimator takes the weighted average of these
two estimators
and so takes the form
%
%e3.1 ###
\begin{eqnarray}\label{genShrink}
\mathbf{f}^*(\omega) = W_T(\omega)\widetilde{\mathbf{V}}(\omega)+
\bigl(1-W_T(\omega
)\bigr)\widetilde{\mathbf{f}}(\omega).
\end{eqnarray}
Our procedure will use the class of vector autoregressive (VAR) models
whose order is selected by the Bayesian information criterion (BIC) for
the parametric estimator and computes the nonparametric estimator with
smoothing spans selected from a plug-in unbiased risk estimation
criterion. The weight at a particular frequency is estimated
data-adaptively so that a~heavier weight is given to the estimator that
gives a better fit at that frequency. We first describe the three
components of the generalized shrinkage estimator before describing the
estimation procedure.

%s3.1 ###
\subsection{Component 1: The parametric estimator}\label{param}
%s3.1.1 ###
\subsubsection{The vector autoregressive process}
Here, we use the class of vector autoregressive models for obtaining a
parametric estimator for the spectral density matrix.
A multivariate time series $\mathbf{X}(t)$ has a VAR($K$)
representation if
$\mathbf{X}(t) = \sum_{k = 1}^K \Phi_{k} \mathbf{X}(t-k) + \mathbf{Z}(t)$,
where the $\Phi_k$'s are $P\times P$ coefficient matrices, $\mathbf
{Z}(t)$ is white noise with covariance matrix $\Sigma_Z$, and other
regularity conditions are satisfied [\citet{Brockwell98}]. The spectral
density matrix for the VAR($K$) time series takes the form
\begin{eqnarray} \label{Eq:VARspec}
\mathbf{V}(\omega) &=& \frac{1}{2\pi}\bigl\{\bigl(\mathbf{I}_{P} - \Phi
_{1}\exp(-i\omega1) - \cdots- \Phi_{K}\exp(-i\omega K)
\bigr)\bigr\}^{-1} \times\Sigma_Z\nonumber
\\[-8pt]\\[-8pt]
&&{}\times \bigl\{\bigl(\mathbf{I}_{P} - \Phi_{1}\exp(-i\omega1) - \cdots-
\Phi_{K}\exp(-i\omega K)\bigr)^*\bigr\}^{-1},\nonumber
\end{eqnarray}
where $\mathbf{I}_{P}$ is the $P\times P$ identity matrix and $^*$
denotes the
complex conjugate transpose. The order $K$ can be determined using some
model selection criterion (such as BIC).

%s3.1.2 ###
\subsubsection{$N$-trial least squares estimation}
We now describe how to obtain the least squares estimates of the
coefficients of a VAR($K$) model for a~multivariate time series
recorded from $N$ trials. The time series for the $n$th trial is
modeled as
%
%e3.2 ###
\begin{eqnarray}\label{VarK}
\mathbf{X}_n(t) = \sum_{k = 1}^K \Phi_{k} \mathbf{X}_n(t-k) +
\mathbf{Z}_n(t),
\end{eqnarray}
where $\mathbf{X}_n(t) \!=\! (X_{1n}(t), \ldots, X_{Pn}(t))^\top\!$,
$\mathbf{Z}_n(t) \!=\! (Z_{1n}(t), \ldots, Z_{Pn}(t))^\top\!$, $t = 1,
\ldots,\break T$, and $n = 1,\ldots,N$. We extend the estimation for a
single-trial\vadjust{\goodbreak} multivariate time series in \citet{Lutkepohl93}.
Suppose we have $K$ many presample values $\mathbf{X}_n(-K+1),\ldots
,\mathbf{X}_n(0)$ for each trial of this time series. For each trial
$n = 1, \ldots N$, define
%
%e3.7 ###
%e3.6 ###
%e3.5 ###
%e3.4 ###
%e3.3 ###
\begin{eqnarray}
\mathbf{X}_n &=& (\mathbf{X}_n(1), \ldots, \mathbf{X}_n(T)) \qquad (P \times T),
\\
\mathbf{B} &=& (\Phi_1, \ldots, \Phi_K) \qquad (P \times PK),
\\
\mathbf{Y}_{n,t} &=&
\pmatrix{
\mathbf{X}_n(t) \cr \vdots\cr \mathbf{X}_n(t-K+1)
}
\qquad (PK \times1),
\\
\mathbf{Y}_n &=& (\mathbf{Y}_{n,0}, \ldots, \mathbf{Y}_{n,T-1}) \qquad
(PK \times T),
\\
\mathbf{Z}_n &=& (\mathbf{Z}_n(1),\ldots,\mathbf{Z}_n(T)) \qquad (P
\times T).
\end{eqnarray}
Now let $b_k^\top$ be the $k$th row of $\mathbf{B}$ and $\mathbf
{X}_{n,(k)} = (\mathbf{X}_{kn}(1), \ldots, \mathbf{X}_{kn}(T))^\top
$ and $\mathbf{Z}_{n,(k)} = (Z_{kn}(1),\ldots,Z_{kn}(T))^\top$. With
this notation, the $n$th trial VAR($K$) model given by equation \eqref
{VarK} can be written as $\mathbf{X}_{n,(k)} = \mathbf{Y}^\top_n
b_k + \mathbf{Z}_{n,(k)}$. \citet{Lutkepohl93} gave the solution in
case $N = 1$ and he showed that the solution is equivalent to OLS
estimation. Now for $N > 1$, the setting becomes analogous to that of
repeatedly measured multivariate data. From this perspective, one can
see that the least squares estimator for the VAR($K$) model with $N$
trials is
$\hat{b}_k = (\sum_{n=1}^N\mathbf{Y}_n\mathbf{Y}_n^\top
)^{-1}(\sum_{n=1}^N\mathbf{Y}_n\mathbf{X}_{n,(k)}).$
Suppose that $NT \gg PK$. Our estimate of $\Sigma_Z = \mathbb
{E}(\mathbf{Z}(t)\mathbf{Z}(t)^\top)$ is
%
%e3.8 ###
\begin{eqnarray}
\widehat{\Sigma}_Z = \frac{1}{NT-PK}\sum_{n=1}^N\{(\mathbf
{X}_n - \widehat{\mathbf{B}}\mathbf{Y}_n)(\mathbf{X}_n - \widehat
{\mathbf{B}}\mathbf{Y}_n)^\top\}.
\end{eqnarray}
The degrees of freedom adjustment is due to the $PK$ many coefficients
in each of the $P$ equations in equation \eqref{VarK}.

Note that in equation \eqref{VarK}, there are $KP^2 + P(P+1)/2$
parameters to estimate. However, it is not unusual for the number of
components of the time series to be large and the number of time points
per trial to be small. Therefore, it is not efficient to estimate the
parameters using only a single trial. Here, we have developed a method
for estimating the parameters by pooling the data over all of the trials.
This is valid if one assumes that the data from each trial is a
realization of a common underlying $\operatorname{VAR}(K)$ process.

The least squares estimation procedure above requires the VAR order $K$
to be known. To select the order $K$ in an $N$-trial multivariate time
series framework, we use the information criterion function
$\operatorname{IC}(\kappa) = \log|\widehat{\Sigma}_Z(\kappa)| + \operatorname
{Pen}(T,N,P,\kappa)$,
where $\widehat{\Sigma}_Z(\kappa)$ is the estimate of $\Sigma_Z$
obtained after fitting a~VAR($\kappa$) model and $\operatorname
{Pen}(T,N,P,\kappa)$ is some penalty function for complexity. Here, we
use the penalty function in BIC, which is $\operatorname{Pen}(T,N,P,\kappa) =
\frac{\log(NT)}{NT}\kappa P^2.$
The optimal order $K$ is selected so that $K = \arg\min_\kappa\operatorname
{IC}(\kappa)$.\vadjust{\goodbreak}

%s3.2 ###
\subsection{Component 2: The nonparametric estimator}\label{nonparam}
%s3.2.1 ###
\subsubsection{\texorpdfstring{$\!\!$The smoothed periodogram matrix}{The smoothed periodogram matrix}}
$\!\!$Let $\mathbf{d}_{\mathbf{X},n}(\omega) \!=\! \frac{1}{\sqrt{2 \pi
}}\sum_{t=1}^T \mathbf{X}_n(t)\times \exp(-i \omega t)$ be the discrete
Fourier transform of $\mathbf{X}_n(t)$. The $n$th trial raw
periodogram is defined to be
$\mathbf{I}_n(\omega) = \frac{1}{T}\mathbf{d}_{\mathbf
{X},n}(\omega)\mathbf{d}^*_{\mathbf{X},n}(\omega)$.
Kernel smoothing is a common method for estimating the spectral density
matrix. Let $w_T(\alpha)$ be a kernel (weight) function
that has smoothing span $M_T$ such that, as $T \rightarrow\infty$,
$M_T \rightarrow\infty$ but $M_T/T \rightarrow0$.
We compute the $n$th trial estimate of the $(j,k)$th element of
$\mathbf{f}(\omega)$ with $\widetilde{f}_{n,jk}(\omega) = \int
_{-\pi}^\pi w_T(\omega- \alpha)I_{n,jk}(\alpha)\,d\alpha$. Under
regularity conditions [\citet{Brillinger01}], this estimate is an
element-wise consistent estimator for $\mathbf{f}(\omega)$. The final
estimate for
$\mathbf{f}(\omega)$ using all of the trials is the average over the
trials, namely,
$\widetilde{\mathbf{f}}(\omega) = N^{-1}\sum_{n=1}^N \widetilde
{\mathbf{f}}_{n}(\omega)$.
Similarly, the elements of $\widetilde{\mathbf{f}}(\omega)$ are
consistent for the elements of ${\mathbf{f}}(\omega)$. Note that in
this setup, we can apply minimal smoothing per trial because the final
nonparametric estimator $\widetilde{\mathbf{f}}(\omega)$ undergoes
further smoothing due to
the averaging across replicated trials.

%s3.2.2 ###
\subsubsection{Automatic selection of the optimal smoothing span}\label{optimalBW}
When $M_T$ is too small, the resulting estimate can capture very
localized peaks but may be too erratic. Conversely,
if it is too large, the resulting estimate will be very smooth but may
miss vital peaks that characterize the process.
Optimal smoothing spans have been studied for univariate time series.
For example, the approach by \citet{Ombao01} used
the full likelihood by minimizing the gamma deviance.
Here, inspired by \citet{Lee97} and \citet{Lee01}, we
develop an
approach for span selection based on the quadratic risk
function.

Define the integrated risk function for a smoothing span $h$ to be
$R(h) = \int_0^\pi\mathbb{E}(\|\mathbf{f}(\omega)- \widetilde
{\mathbf
{f}}_{h}(\omega)\|^2)\,d\omega$,
where $\widetilde{\mathbf{f}}_h(\omega)$ is the periodogram matrix
smoothed with a kernel having span size $h$ and $\|\cdot\|^2$ is the
normalized Hilbert--Schmidt norm defined by $\|A\|^2 = P^{-1}\operatorname
{tr}(AA^*)$. The goal is to pick $M_T$ so that $M_T = \arg\min
_hR(h)$. Such an $M_T$ is the global optimal smoothing span. However,
we cannot evaluate $R(h)$ because $\mathbf{f}(\omega)$ is unknown. An
approach in the
univariate time series context is to use
a plug-in unbiased risk estimation (PURE) procedure by plugging in a
pilot estimator $\widetilde{\mathbf{f}}_{\mathrm{pilot}}(\omega)$ for
$\mathbf{f}(\omega)$,
where the pilot estimator is the periodogram smoothed by a kernel with
an arbitrarily picked smoothing span. This gives an asymptotically
unbiased estimate of $R(h)$.

The proposed procedure for obtaining the optimal smoothing span for the
$n$th trial of a multivariate time series is as
follows. We combine the PURE procedure with a leave-one-out procedure.
Let $\widetilde{\mathbf{f}}_{n,h}(\omega)$ be the $n$th trial
periodogram smoothed with smoothing span $h$. Let $\hat{\mathbf
{f}}_{(-n)}(\omega) = (N-1)^{-1}\sum_{j=1, j\neq n}^N \mathbf
{I}_j(\omega)$. Since\vspace*{1pt} each $\mathbf{I}_j(\omega)$ is an
approximately unbiased estimator of $\mathbf{f}(\omega)$, then
$\hat{\mathbf
{f}}_{(-n)}(\omega)$ is an unbiased estimate of $\mathbf{f}(\omega)$
and thus will
serve as the pilot estimator. Then the $n$th trial integrated PURE is
$\widehat{R}_n(h) = \int_0^\pi\|\hat{\mathbf{f}}_{(-n)}(\omega
) - \widetilde{\mathbf{f}}_{n,h}(\omega)\|^2 \,d\omega$.
We proceed by picking the smoothing span $M^{(n)}_T$ for the $n$th
trial so that $M^{(n)}_T = \arg\min_h \widehat{R}_n(h)$.

%s3.3 ###
\subsection{Component 3: The theoretical shrinkage weight}
Ideally, the frequen\-cy-specific shrinkage weight corresponding to the
parametric estimator, denoted $W_T(\omega)$, should be greater than
0.5 at frequencies where the parametric estimate gives a better fit
than the nonparametric estimate. In fact, as we will see later, the
shrinkage weight is a function of the mean squared error of each of the
parametric and nonparametric estimators. We formalize this in the
following discussion.

First, we introduce the notation
$\mathbf{f}_T^{\,0}(\omega) = \mathbb{E}(\widetilde{\mathbf
{f}}_T(\omega))$, where $\widetilde{\mathbf{f}}_T(\omega)$ is the
smoothed periodogram. For ease, we drop the subscript $T$ in
the notation. From \citet{Brillinger01}, $\mathbf{f}^{\,0}(\omega) -
\mathbf{f}(\omega) \rightarrow\mathbf{0}$ as $T\rightarrow\infty
$. Here, $\mathbf{f}^{\,0}(\omega)$ will serve as the proxy for $\mathbf
{f}(\omega)$ and will be utilized in deriving the weights for the
generalized shrinkage estimator. Define the squared error loss function
$\mathcal{L}(\mathbf{f}^*(\omega), \mathbf{f}^{\,0}(\omega)) =
\|\mathbf{f}^*(\omega
)-\mathbf{f}^{\,0}(\omega)\|^2,$ where the norm is the normalized
Hilbert--Schmidt norm.
Then the risk function is
$\mathcal{R}(\mathbf{f}^*(\omega), \mathbf{f}^{\,0}(\omega)) = \mathbb{E}
(\mathcal{L}(\mathbf{f}^*(\omega), \mathbf{f}^{\,0}(\omega)) )$.
The optimal shrinkage weight $W_T(\omega)$ is the minimizer of the
risk function.
First, define
%
%e3.10 ###
%e3.9 ###
\begin{eqnarray}\label{alpha2}
\alpha_T^2(\omega) &=& \operatorname{MSE}(\widetilde{\mathbf{V}}(\omega))
= \mathbb{E}\bigl(\|\widetilde{\bf V}(\omega) - \mathbf{f}^{\,0}(\omega)\|^2\bigr),
\\
\label{beta2}
\beta_T^2(\omega) &=& \operatorname{MSE}(\widetilde{\mathbf{f}}(\omega)) =
\mathbb{E}\bigl(\|\widetilde{\mathbf{f}}(\omega)- \mathbf{f}^{\,0}(\omega
)\|^2\bigr)
\end{eqnarray}
and
%e3.11 ###
\begin{eqnarray}
\label{delta2}
\delta_T^2(\omega) = \mathbb{E}\bigl(\|\widetilde{\bf V}(\omega)
- \widetilde{\mathbf{f}}(\omega)\|^2\bigr).
\end{eqnarray}
Then it can be easily shown the optimal shrinkage weight is
%
%e3.12 ###
\begin{equation}\label{optimalWeight}
W_T(\omega) = \frac{\beta_T^2(\omega)-0.5(\alpha_T^2(\omega
)+\beta_T^2(\omega)-\delta_T^2(\omega))}{\delta_T^2(\omega)}.
\end{equation}
This result generalizes that given by both \citet{Bohm08},\vadjust{\eject} whose
shrinkage target was that of a one-factor model, and \citet{Bohm09},
whose shrinkage target was the scaled identity matrix.

\begin{remarks*} Upon inspection of the optimal shrinkage
weight equa-\break tion~\eqref{optimalWeight}, one can obtain insight that the
generalized shrinkage estimator behaves analogous to the Bayes
estimators, where the prior estimator is given by the parametric
estimator and the data-driven estimator is given by the nonparametric
estimator. Note that the behavior of the shrinkage weight is a function
of the relative performance of each of the parametric and nonparametric
estimators. In particular, if the parametric estimator models the
spectral density matrix well so that $\alpha_T^2(\omega) \rightarrow
0$ at a rate much faster than $\beta_T^2(\omega) \rightarrow0$, then
the shrinkage weight will shift toward the parametric estimator;
otherwise, the shrinkage weight will shift toward the nonparametric
estimator. The second term of the numerator of equation \eqref
{optimalWeight} corrects for the correlation between the parametric
estimator and the nonparametric estimator. Empirical Bayes estimators
use the data to construct the prior estimator. Our approach is
analagous in the sense that we use the same data to construct the
parametric and nonparametric estimators, and so the second term takes
into account that the two estimators are highly likely to be
correlated. Back to equation~\eqref{optimalWeight}, we see that the
denominator makes the generalized shrinkage estimator robust to the
misspecification of the parametric estimator. The denominator is the
squared distance between the parametric and nonparametric estimators.
So if the parametric and nonparametric estimators are vastly different
from each other, then the denominator of the weight will be large, and
so the weight will be larger for the nonparametric estimator.
\end{remarks*}

%s3.4 ###
\subsection{Constructing the generalized shrinkage estimator}\label{constructGenV}

The parametric and nonparametric components of the generalized
shrinkage estimator can be obtained using standard procedures.
To estimate the shrinkage weight, we need to construct an estimate of
each of $\alpha^2_T(\omega)$ and $\beta^2_T(\omega)$, and $\delta
^2_T(\omega)$.

Each of $\alpha^2_T(\omega)$ and $\beta_T^2(\omega)$ is the
expected distance from their respective estimator with $\mathbf
{f}^{\,0}(\omega)$, and so
we need to provide an estimate of $\mathbf{f}^{\,0}(\omega)$. An
asymptotically unbiased
estimator of $\mathbf{f}^{\,0}(\omega)$ is the average of\vspace*{1pt} the periodograms:
$\hat{\mathbf{f}}^{\,0}(\omega)= N^{-1}\sum_{n=1}^N\mathbf
{I}_n(\omega)$.
Consider $\beta_T^2(\omega)$. Since $\widetilde{\mathbf{f}}(\omega
)$ is an unbiased
estimator of $\mathbf{f}^{\,0}(\omega)$ by construction, then $\beta
_T^2(\omega)$ is the
sum over the variances of each of the elements of $\widetilde{\mathbf
{f}}(\omega)$. Assuming
that each element of~$\hat{\mathbf{f}}^{\,0}(\omega)$ varies slowly
over frequency, our
proposed estimator is a type of sample variance, that is, we look at
the window of size $C_T$ around $\widetilde{\mathbf{f}}(\omega)$ for
each~$\omega$:
%
%e3.13 ###
\begin{eqnarray}\label{betahat}
\hat{\beta}^2_T(\omega) = C_T^{-1}\sum_{k =
-(C_T-1)/2}^{(C_T-1)/2}\| \widetilde{\mathbf{f}}(\omega)- \hat
{\mathbf{f}}^{\,0}(\omega+
\omega_k)\|^2.
\end{eqnarray}
Our simulation studies indicate the choice of this procedure is robust
over a wide choice for $C_T$. We estimate $\alpha_T^2(\omega)$ using
a type of plug-in estimator smoothed over frequencies:
%
%e3.14 ###
\begin{eqnarray}\label{alphahat}
\hat{\alpha}^2_T(\omega) = C_T^{-1}\sum_{k =
-(C_T-1)/2}^{(C_T-1)/2}\|\widetilde{\mathbf{V}}(\omega)- \hat
{\mathbf{f}}^{\,0}(\omega+
\omega_k)\|^2.
\end{eqnarray}
Our proposed estimator for $\delta_T^2(\omega)$ is constructed in a
similar manner:
\begin{eqnarray}\label{deltahat}
\hat{\delta}^2_T(\omega) &=& \frac{1}{2}\Biggl[C_T^{-1}\sum_{k =
-(C_T-1)/2}^{(C_T-1)/2}\bigl(\|\widetilde{\mathbf{f}}(\omega+
\omega_k) - \widetilde{\mathbf{V}}(\omega)\|^2\nonumber
\\[-8pt]\\[-8pt]
&&{}\hspace*{91pt}+ \|\widetilde
{\mathbf{V}}(\omega+ \omega
_k) - \widetilde{\mathbf{f}}(\omega)\|^2\bigr)\Biggr].\nonumber
\end{eqnarray}
Then we can obtain a plug-in estimate of the optimal shrinkage weight using
%
%e3.15 ###
\begin{eqnarray}
\widehat{W}_T(\omega) = \frac{\hat{\beta}^2_T(\omega
)-0.5(\hat{\alpha}_T^2(\omega) + \hat{\beta}_T^2(\omega)
- \hat{\delta}_T^2(\omega))}{\hat{\delta}^2_T(\omega)}.
\end{eqnarray}
Note that due to estimation error in obtaining $\hat{\alpha
}_T^2(\omega)$, $\hat{\beta}_T^2(\omega)$ and $\hat
{\delta}_T^2(\omega)$, it is possible for our estimate of the
shrinkage weight to fall outside the interval $[0,1]$. If this occurs,
we truncate the estimated weight to 0 or 1. Finally, we plug in the
estimated weight to obtain an estimate of the generalized shrinkage estimator:
%
%e3.16 ###
\begin{eqnarray}\label{shrinkageEst}
\hat{\mathbf{f}}^*(\omega) = \widehat{W}_T(\omega)\widetilde
{\mathbf{V}}(\omega)+
\bigl(1-\widehat{W}_T(\omega)\bigr)\widetilde{\mathbf{f}}(\omega).
\end{eqnarray}

%s4 ###
\section{Simulation study}\label{Simulations}
We now compare the performance of the generalized shrinkage estimator
against competitors which
are, namely, the VAR estimator whose order is selected using the BIC
criteria, the smoothed periodogram and the multitaper [\citet
{Percival93}]. The VAR estimator was estimated as described in Section
\ref{param}. The smoothed periodogram, as described in Section \ref
{nonparam}, is obtained using the Hann kernel whose smoothing span is
objectively determined by the quadratic risk criterion
described in Section \ref{optimalBW}. The multitaper estimator was
obtained by averaging the per-trial multitaper estimates, which is
similar to how we obtained the smoothed periodogram estimator. The
optimal number of tapers was also objectively determined using
a PURE procedure similar to that described in Section~\ref{nonparam}.
The estimators are compared using the mean squared error (MSE).

The true underlying process was a sum of a VAR($5$) process and a
first-order vector moving average (VMA) process. The two processes were
generated independently so that the underlying spectral density matrix
is the sum of that of the VAR and the VMA. The vector time series had
$P = 12$ dimensions, $N = 120$ trials, and $T = 256$ time points. The
values of the parameters are given in the \hyperref[app]{Appendix}.

%f1 ###
\begin{figure} % float placement: (h)ere, page (t)op, page (b)ottom, other (p)age
\tabcolsep=0pt
\begin{tabular}{c}

\includegraphics{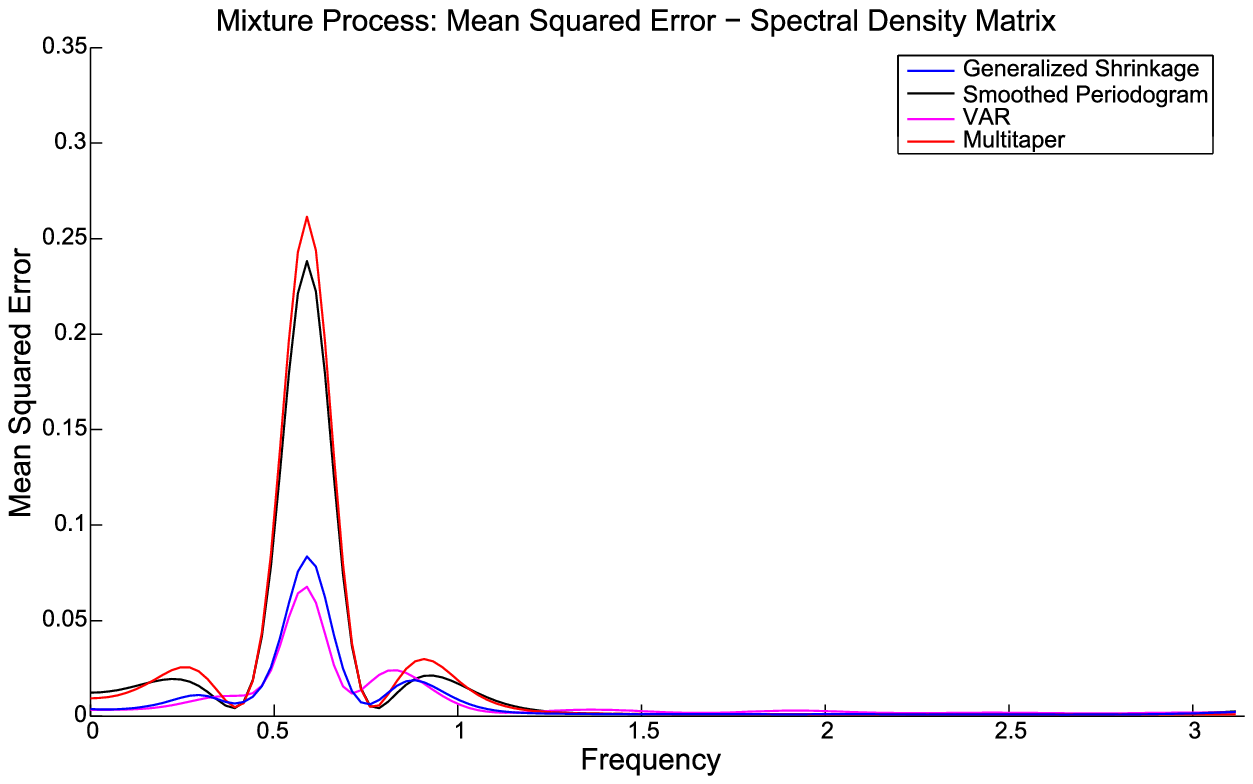}
\\
  (a)\\[6pt]

\includegraphics{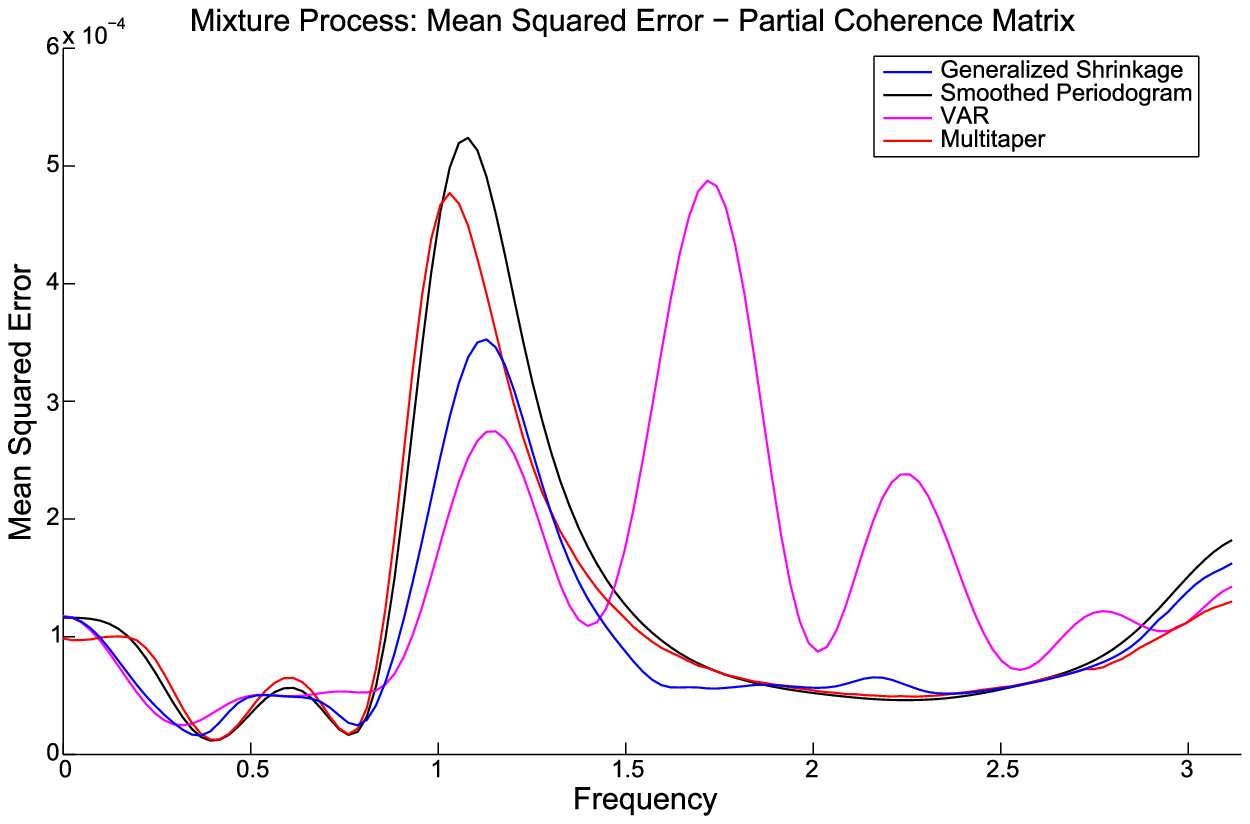}
\\
  (b)
  \end{tabular}
\caption{Mean squared error estimated via Monte Carlo for the simulation.
\textup{(a)} Frequency-specific mean squared error of each estimate of the spectral density matrix.
\textup{(b)} Frequency-specific mean squared error of each estimate of the partial coherence matrix.}
\label{varVma}
\end{figure}

%f2 ###
\begin{figure}

\includegraphics{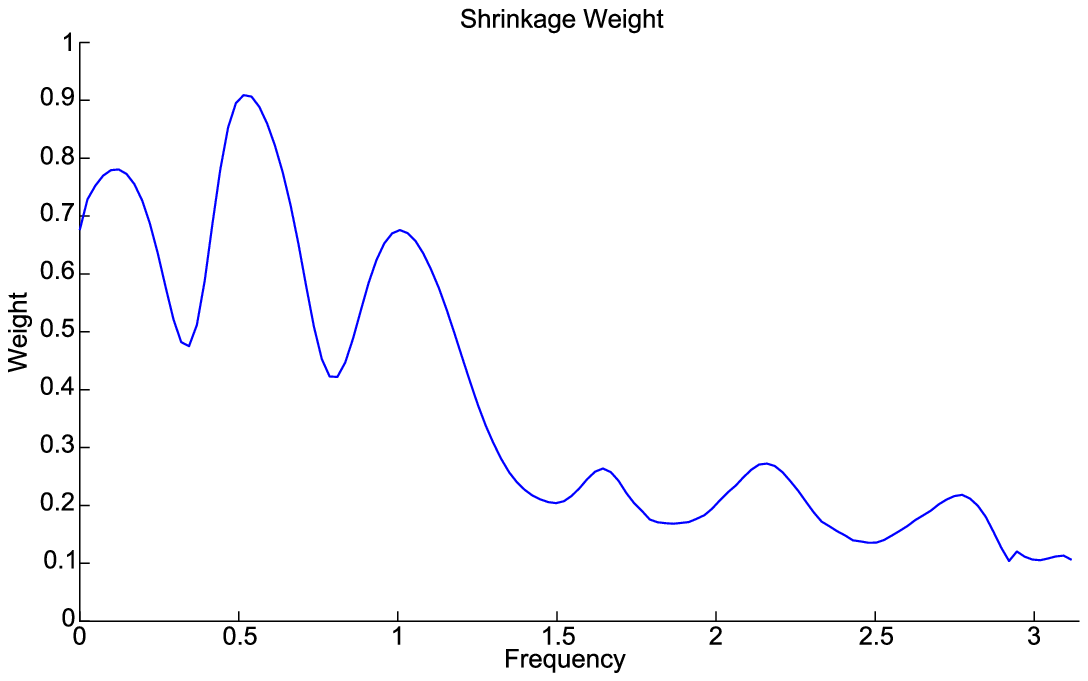}

  \caption{Shrinkage weight for the generalized shinkage estimator estimated via Monte Carlo for the simulation.}
  \label{simWeight}
\end{figure}

The results of the simulation study are shown in Figure \ref{varVma}.
Under this setting, it is obvious that the VAR parametric estimator was
incorrectly specified. However, as just noted earlier, a VAR can
capture the peaks in the spectral density matrix.
Note that while the VAR estimator is the best estimator of the spectral
density matrix in the sense that it accurately captured the peaks in
the autospectra, it was not necessarily the best estimator of partial
coherence (because the latter is a highly nonlinear function of the
former). This can be seen in the high frequencies as shown in Figure
\ref{varVma}. In other words, the best estimator of the spectral
density matrix is not necessarily the best estimator for partial coherence.
The nonparametric estimators performed poorly relative to the
parametric VAR estimator in estimating the spectral density matrix
because each oversmoothed the peaks in the autospectra.
The generalized shrinkage estimator performed well in estimating both
the spectral density matrix and partial coherence. It is
clear that the generalized shrinkage estimator borrowed strength from
the parametric VAR estimator to better capture the peaks in the
autospectra so that it estimated the spectral density matrix much
better than the nonparametric estimators.
This can be seen by the shrinkage weight, as shown in Figure \ref
{simWeight}. The shrinkage weight is near 1.0 at frequencies near the
location of the peaks of the autospectra.

%s5 ###
\section{Functional connectivity for EEG}\label{EEG}
The study of functional connectivity of brain signals is the
investigation of the dependencies between brain signals that have been
measured from spatially separated regions of the brain [\citet
{Friston93}]. We investigated the dependency structure in the EEG
signals between certain regions and how it differs across experimental
conditions in a visual-motor task.

%s5.1 ###
\subsection{Data description and preprocessing}
The EEG data were recorded from the scalp using a 64-channel EEG system
(EMS, Biomed, Korneuburg, Germany). The electrodes were applied to the
scalp using the standard International 10--20 system with a reference
electrode on the nose-tip. The EEG signals were recorded at 512
samples/second/channel and filtered using a~high-pass filter of 0.02 Hz
and a low-pass filter of 100 Hz.

\begin{figure} % float placement: (h)ere, page (t)op, page (b)ottom, other (p)age

\includegraphics{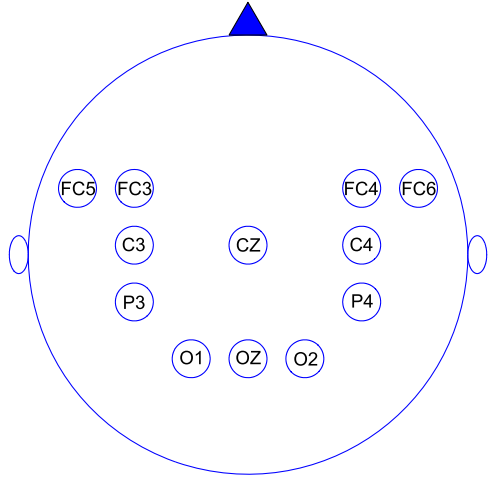}
\vspace*{-5pt}
  \caption{Relative placement of the twelve EEG channels preselected for the analysis.  These channels were presumed to be recording the relevant neural processes for the visual-motor task in the experiment.}
  \label{scalp}\vspace*{-6pt}
\end{figure}

Participants of the study were required to make quick displacements of
a~hand-held joystick from a central position either to the right or to
the left from center as instructed by a visual cue. The visual cue
randomly selected the movement per trial. From a standard montage of 64
scalp electrodes, our neuroscientist collaborator selected a subset of
$P = 12$ channels that were presumed, based on published studies, to be
recording the relevant neural processes for these visual-motor actions
[\citet{Marconi01}; \citet{Bedard09}]. These electrode
sites include the
fronto-central leads (FC) to measure activity related to premotor
processing, the central leads (C) to measure activity related to motor
performance, and the parietal~(P) and occipital (O) leads to measure
activity related to visual-motor transformations. We did not add more
electrodes in the analysis because this will present unnecessary
computing and modeling complications, which we will later discuss. The
relative locations of these 12 electrode sites are shown in Figure \ref{scalp}.

\begin{figure}[b] % float placement: (h)ere, page (t)op, page (b)ottom, other (p)age
\vspace*{-6pt}
\includegraphics{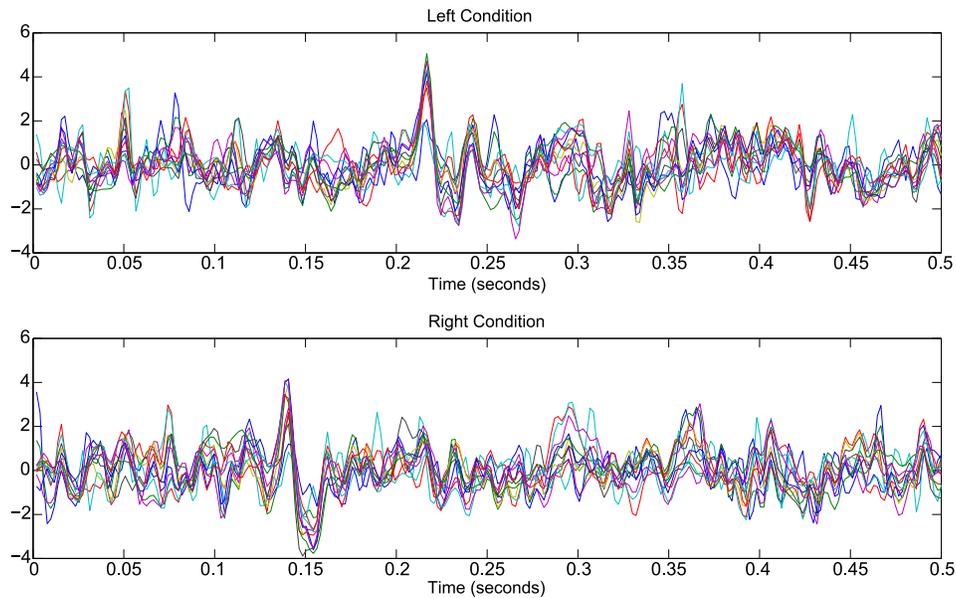}%
\vspace*{-5pt}
  \caption{Representative 12-channel filtered and detrended EEG recorded from one trial for each of the {\it left} and {\it right} conditions.}
  \label{SingleTrialTS}
\end{figure}

The methods that we have developed in this work have been for
single-subject analyses, and so here, we show results for only one
subject. We analyze the first 0.5~seconds from stimulus onset of the
EEG signals, yielding a multivariate time series having length $T =
256$ per trial, and there were $N_{\mathrm{L}} = 118$ and $N_{\mathrm{R}}
= 138$ trials for the leftward and rightward movements respectively.
For the analysis, we removed the linear and quadratic trends from the
subject's EEGs. The EEGs were further filtered using a $4$th-order
low-pass Butterworth filter with stopband at 50~Hz and then
standardized to have unit variance. Figure \ref{SingleTrialTS}
illustrates time plots of the $P = 12$ filtered and detrended EEG
signals obtained from a representative participant during leftward and
rightward joystick movements.

%4
%f3 ###

%3
%f4 ###

%s5.2 ###
\subsection{Details on the statistical procedure}
\begin{figure}

\includegraphics{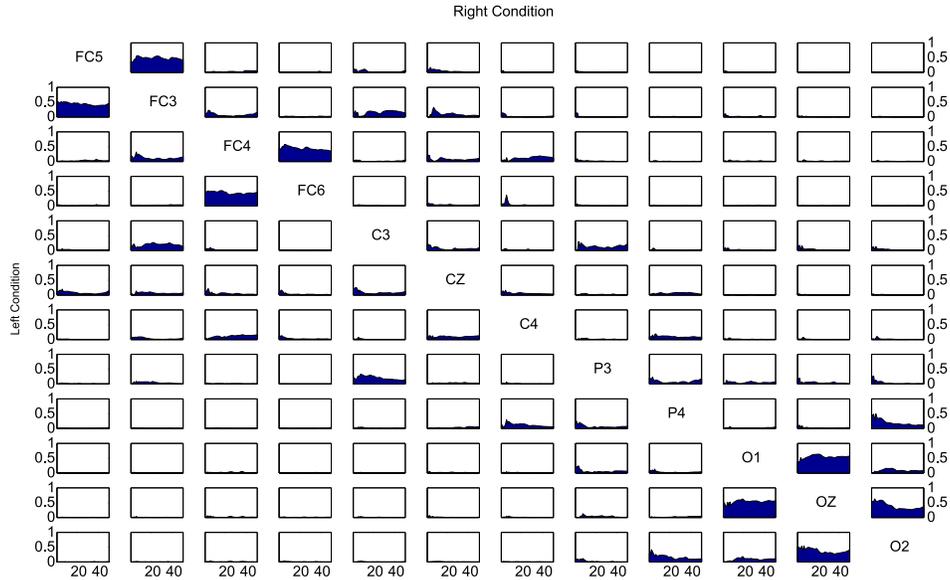}

  \caption{Estimated partial coherence for all pairs of the twelve channels in the analysis for the ``right'' condition (upper triangle) and the ``left'' condition (lower triangle).}
  \label{PCoh}
\end{figure}

\begin{figure}

\includegraphics{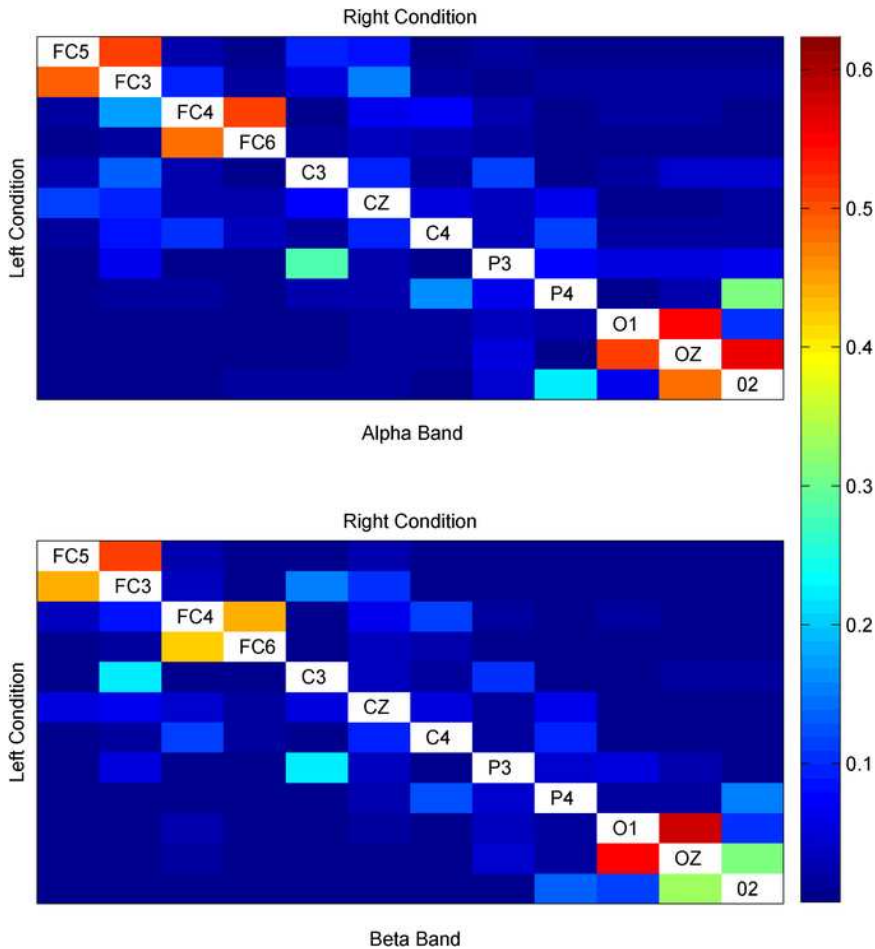}

  \caption{Relative strength of functional connectivity measured by partial coherence at the alpha and beta frequency bands for the ``right'' condition (upper triangle) and ``left'' condition (lower triangle).}
  \label{LR}
\end{figure}

Our aim in this work was two-fold: first, to estimate the strength of
functional connectivity as measured by partial coherence, and second,
to identify which connections differentiate between the ``left'' and
the ``right'' trials. The parametric component was computed using a
VAR(19) model where the order was selected using the BIC criterion and
the parameters were estimated using the $N$-trials least squares
procedure. The nonparametric component was obtained using the Hann
kernel with smoothing span automatically selected by our procedure. The
mean smoothing span selected for the trials for the ``left'' conditions
was 23.37 with standard deviation 7.74 and for the ``right'' conditions
22.04 with standard deviation 7.75.
The estimated partial coherences are shown in Figure \ref{PCoh}. To
perform a frequency-band analysis of functional connectivity, we
computed the partial coherences averaged over the frequencies in the
band of interest. The partial coherences for each of the alpha band
(8--12~Hz) and beta band (18--30 Hz) for each of the ``left'' and
``right'' conditions are shown in Figure \ref{LR}.

%8
%f5 ###

\begin{figure}

\includegraphics{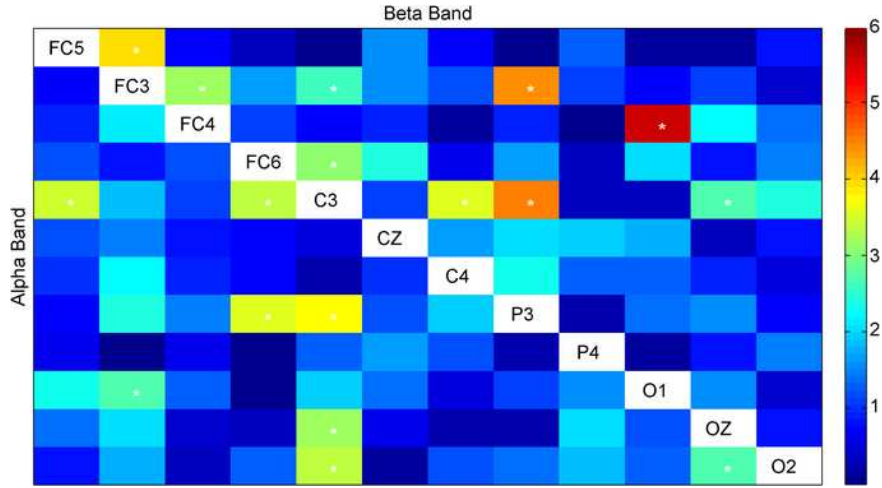}

  \caption{The absolute value of the $t$-statistics for testing for differences between ``left'' and ``right'' conditions.
   The $t$-statistics marked with an asterisk (*) were declared significant after adjusting their $p$-values at FDR level of 0.05.}
  \label{tstats}
\end{figure}

To test for differences in strength of connectivity over a frequency
band, we normalized the partial coherence estimates using the Fisher's
$Z$-transform. We used the jackknife to estimate the standard error of
the estimated partial coherences over the bands. This was done by,
after leaving out one trial, estimating partial coherence via the
generalized shrinkage estimator of the spectral density matrix and then
using Fisher's $Z$-transform to normalize these jackknifed partial
coherence estimates. This allowed us to obtain a~jackknife sample of
size $N_{\mathrm{L}}$ estimates of partial coherence for each of the
frequency bands for the ``left'' condition and a jackknife sample of
size $N_{\mathrm{R}}$ estimates of partial coherence for each of the
frequency bands for the ``right'' condition. The point estimates for
each of the conditions as well as their standard errors, both estimated
using a sample size $N_{\mathrm{L}}$ or $N_{\mathrm{R}}$, allowed us to
create $t$-statistics to test for differences via a two-sample $t$-test
across conditions. These $t$-statistics are shown in Figure \ref
{tstats}. Note that for each frequency band, we performed $12\cdot
(12-1)/2 = 66$ tests, so that in all we are performing 132 tests. To
correct for multiple comparisons, we performed our tests controlling
for FDR at $0.05$. The null hypotheses of no difference across
conditions that were rejected are marked with an asterisk in Figure
\ref{tstats}.

%9
%f6 ###

%10
%f7 ###

%5
%f8 ###
\begin{figure}
\tabcolsep=0pt
  \begin{tabular}{c}

\includegraphics{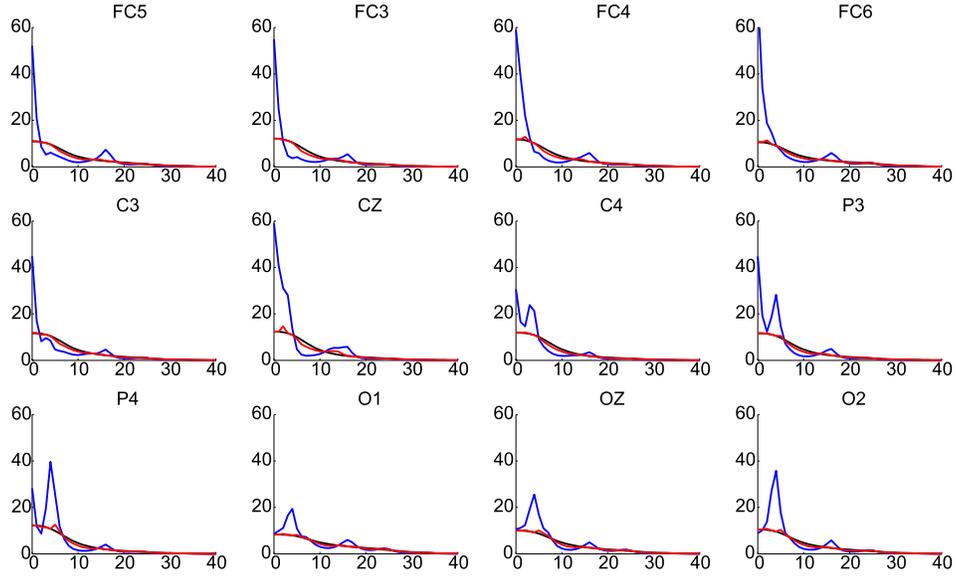}
\\
  (a)\\[6pt]

\includegraphics{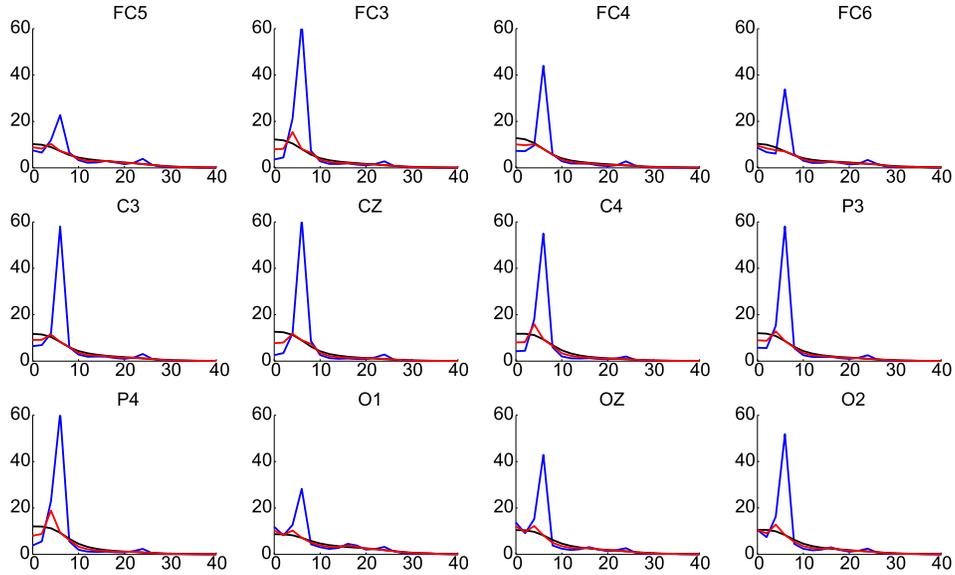}
\\
  (b)
  \end{tabular}
  \caption{Estimated autospectra for each of the two conditions.
  Shown here are the three estimators: the \textup{VAR}(19) (blue), the smoothed periodogram (black), and the generalized shrinkage (red).
  \textup{(a)}~Estimated autospectra for the ``left'' condition.
 \textup{(b)} Estimated autospectra for the ``right'' condition.}\label{psd1}
\end{figure}

%6
%f9 ###
\begin{figure}

\includegraphics{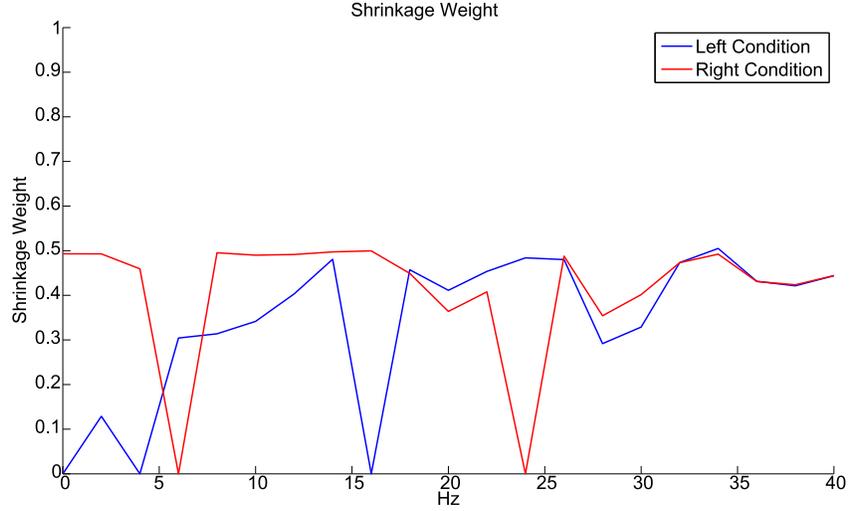}
\vspace*{-5pt}
  \caption{Estimated shrinkage weight.}
  \label{weight}\vspace*{-5pt}
\end{figure}

%7
%f10 ###
\begin{figure}[b]
\vspace*{-5pt}
\includegraphics{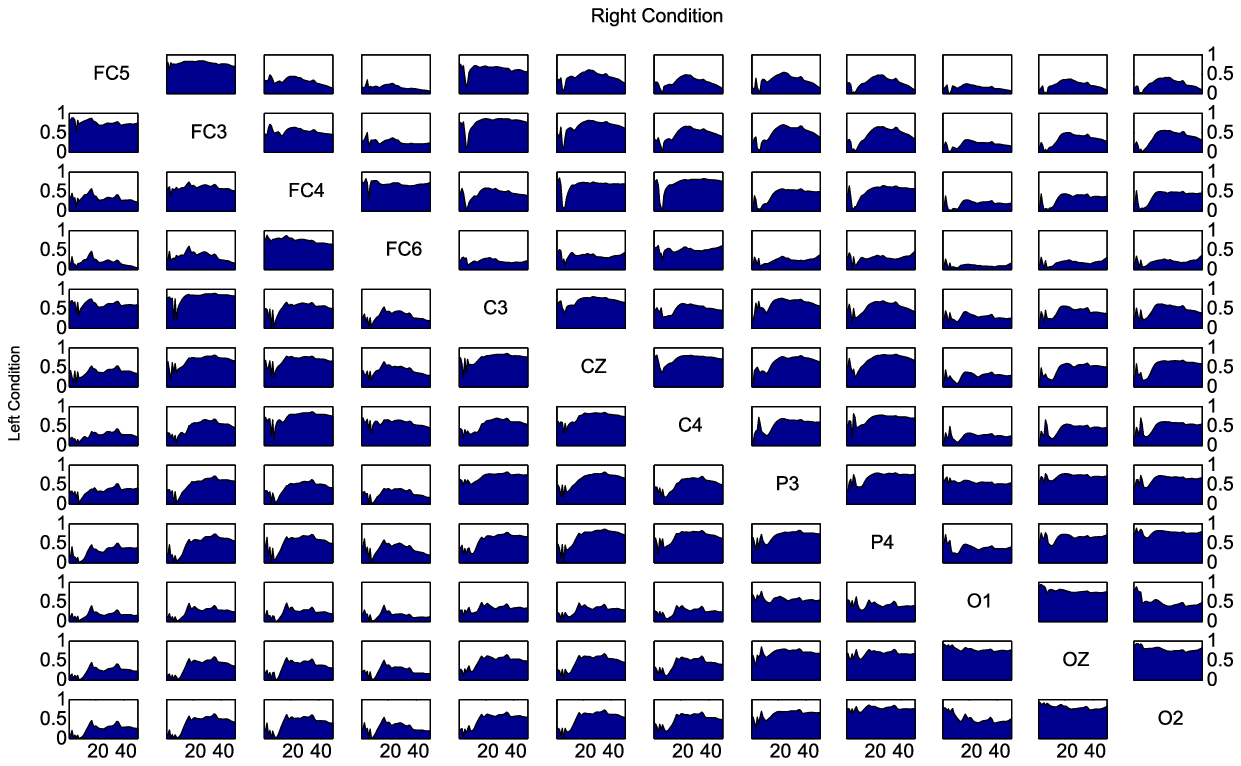}
\vspace*{-5pt}
  \caption{Estimated coherence for all pairs of the twelve channels in the analysis for the ``right'' condition
  (upper triangle) and the ``left'' condition (lower triangle).}
  \label{Coh}
\end{figure}

%s5.3 ###
\subsection{Results}

\textit{On the estimates of the autospectra.} The estimated
power spectral densities for each channel are shown in Figure \ref
{psd1}(a) and (b), and the estimated shrinkage weight is shown in
Figure \ref{weight}. For the ``left'' condition, the VAR estimator
picks up a peak in the autospectra at the very low frequencies, and
then a smaller peak around 16 Hz. At these frequencies, the generalized
shrinkage estimator disagreed with the VAR estimator, and, in fact, the
shrinkage weight was
truncated to $0.0$ at these frequencies. For the ``right'' condition,
again the VAR estimator picks up two peaks, one around 6 Hz and the
other around $24$ Hz, and the smoothed periodogram oversmoothed these
peaks. The estimate of the shrinkage weight was not truncated at 0.0 at
4 Hz, and so the estimates given by the generalized shrinkage estimator
are showing a slight peak. The shrinkage weight was much less than
$0.5$, and so the smoothed periodogram was favored. This implies that
the VAR model may not be an adequate model.

\textit{On the estimates of partial coherence.} Figure \ref
{Coh} shows the estimated coherence for all pairs. Recall that
coherence is the frequency domain analog of squared cross-correlation.
Coherence among the fronto-central leads and among the occipital leads
are strong, in fact, near 1.0 at some frequencies. The coherence
between the occipital leads and the fronto-central leads is smaller.
Since coherence captures both direct and indirect connectivity, then we
can see that the occipital leads are somehow connected with the
fronto-central leads. Partial coherence captures only the direct
connectivity by removing the effects of the other leads in the
analysis. In Figure \ref{PCoh} we see that the strong direct
connections are among the fronto-central leads and the occipital leads.
Moreover, the connections between the fronto-central leads and the
occipital leads are much weaker after partialization, suggesting that
the connection between these regions is more likely to be indirect.

\textit{On comparing connectivity between left vs right
conditions.} When testing for differences of the ``left'' and ``right''
connections as shown in Figure \ref{tstats}, some of the differences
in strength of connections were deemed statistically significant.
However, upon closer inspection, many of these differences may be
considered to be irrelevant because the estimated strength of
connection is very weak. For instance, the difference in strength in
the O1--FC4 connection in the beta band had the largest $t$-statistic
and was considered statistically significant, and yet, the estimated
partial coherence values for this connection are $0.0225$ and $0.0116$
for the ``left'' and ``right'' conditions, respectively. Though this is
nearly a $2$-fold increase in strength of connection, these squared
correlation values are small and so the connection may not be relevant
to the visual-motor task.
Among those pairs where the estimated partial coherence is larger than
0.05 for at least one of the conditions, differences between
conditions were significant for the following pairs: C3--FC5, P3--C3 and
O2--OZ in the alpha band and FC3--FC5, FC3--FC4, C3--FC3 and P3--C3 in the
beta band.

Of these differences, the only connection where it is stronger in the
``left'' condition is the P3--C3 connection; for the other differences,
the ``right'' condition yielded stronger connections. An analysis by
\citet{Bohm10} concluded that the coherence between C3 and FC3
was the
most discriminating feature between the two conditions using
frequency-domain characteristics of the EEG signals. Our jackknife
procedure is consistent with this finding and even provides additional
information, namely, that a measure of the {\it direct} connection
between C3 and FC3 may be used to distinguish between ``left'' and
``right'' conditions.

\textit{On the limitations of EEG.} While EEGs have excellent
temporal resolution, they are not highly localized in space.
Nevertheless, they remain highly utilized in many studies because they
have excellent temporal resolution and thus can be used to probe into
brain processes that occur at the millisecond level. Moreover, they are
relatively inexpensive to collect, noninvasive and highly portable, and
thus have the high potential for brain--computer interface applications.

\textit{On the limitations of partial coherence.}
Partial coherence is a tool for investigating second-order dependencies
for a given set of channels. Partial coherence measures only the
strength of direct dependencies and does not give information of the
direction of the dependencies, which can be captured using other
metrics of association [e.g., \citet{Kaminski01}]. Moreover, certain
regions may be strongly connected but in a nonlinear manner, and if
this is the case, we would have missed this in the present analysis
because partial coherence measures only linear associations between the
signals and the spectral density matrix captures only second-order
dependencies. One can investigate nonlinear and higher order
dependencies in multivariate time series using metrics such as mutual
information.

%f11 ###
\begin{figure}

\includegraphics{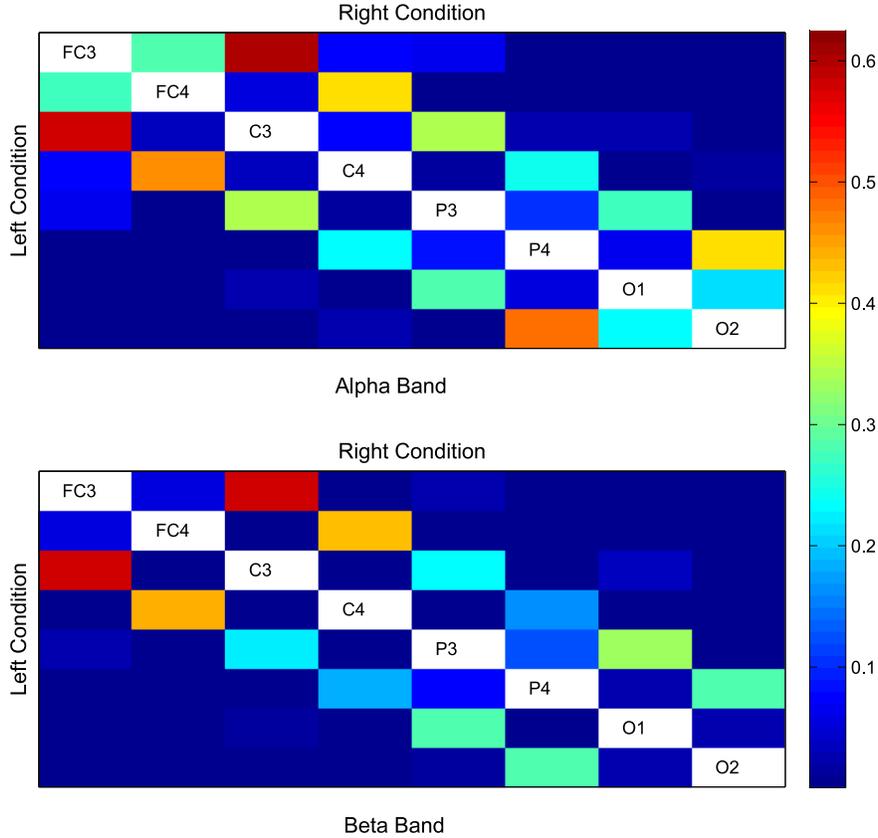}

  \caption{Relative strength of functional connectivity measured by partial coherence at the alpha
  and beta frequency bands for the ``right'' condition (upper triangle) and ``left'' condition
  (lower triangle) when $P = 8$ channels are analyzed.}
  \label{PCoh8}
\end{figure}

%f12 ###
\begin{figure}

\includegraphics{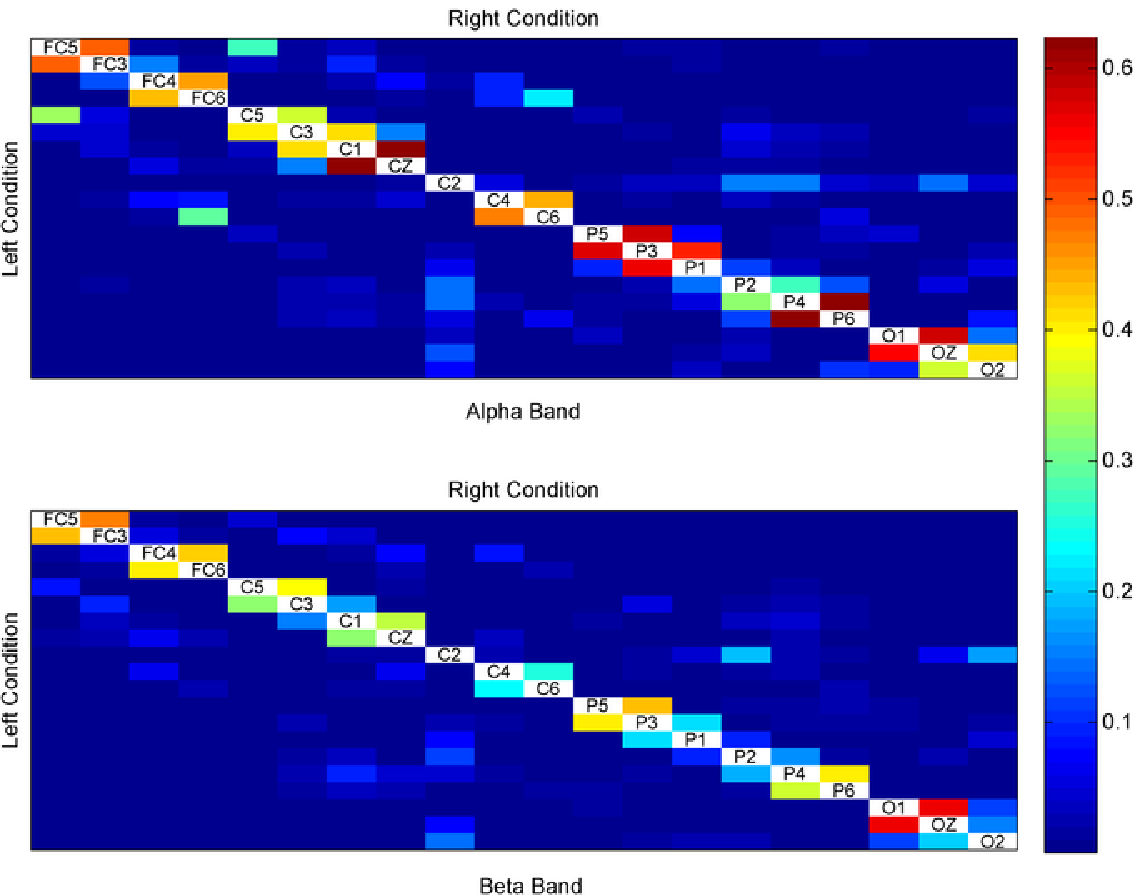}
\vspace*{-5pt}
  \caption{Relative strength of functional connectivity measured by partial coherence at the alpha
  and beta frequency bands for the ``right'' condition (upper triangle) and ``left'' condition
  (lower triangle) when $P = 20$ channels are analyzed.}
  \label{PCoh20}\vspace*{-5pt}
\end{figure}

\textit{On the importance of the selection of the channels.}
Since partial coherence measures direct dependencies, the choice of the
channels to include in the analysis is important. We illustrate this by
showing how the estimates of the partial coherence are affected when
signals from more and less channels are included in the analysis.

We first see the effects if channels were removed. Figure \ref{PCoh8}
shows the estimates of partial coherence for $P = 8$. Compare the
estimates of partial coherence between FC3 and FC4 in the alpha band as
shown in Figure \ref{PCoh8} with that shown in Figure \ref{LR}. The
partial coherence is larger for the case $P = 8$ because the signals
from the FC5 and FC6 channels were excluded. Then in the 12-channel
analysis, when computing partial coherence between FC3 and FC4, the
effects of FC5 were removed from FC4, and since FC5 is highly
conditionally dependent with FC3 (as shown in Figure \ref{LR}), this
dampened the partial coherence between FC3 and FC4. In fact, partial
coherence estimates are larger for the $P = 8$ analysis, and this is
due to the exclusion of certain channels.

Consider now the situation when more channels are included. Figure \ref
{PCoh20} shows the estimates of partial coherence for $P = 20$. Here,
partial coherence estimates are smaller due to the inclusion of certain
channels. Let us look at the partial coherence between P3 and C3 in the
alpha band. Now for $P = 20$, partial coherence between P3 and C3
removes the effects of each of C5 and P5. By removing the effects of P5
from C3, one is also removing the effects of P3 because P3 and P5 are
highly conditionally dependent, and, similarly, by removing the effects
of C5 from P3, one is removing the effects of C3 because C3 and C5 are
highly conditionally dependent. Thus, the partial coherence between P3
and C3 is lower when C5 and P5 are included in the analysis.

This motivates the importance of the choice of channels in the
analysis. As we have just shown, including or excluding channels can
have an effect on the analysis. However, one should not include
everything in the analysis for two reasons. First, there is the problem
of high-dimensionality, which can lead to problems in inference because
there are more connectivity measures to estimate. Second, we note that
many of these (neighboring) channels are highly correlated with each
other, which then introduces redundant information. So if all channels
were included in the analysis, we expect that many of the partial
coherence estimates will be dampened down to 0.

%s6 ###
\section{Discussion}\label{Discussion}

Our main contribution for spectral analysis in multivariate time series
is a new estimator for estimating the spectral density matrix. Our
approach simply takes the weighted average of two known estimators,
namely, a~parametric and nonparametric estimator. We derived
the weights using a multivariate mean squared error criterion. We have
made further developments in shrinkage estimation for the spectral
density matrix by giving freedom in the choice of the shrinkage target
and by giving the appropriate shrinkage weight for this choice of the
shrinkage target. Our shrinkage target in this work is the spectral
density matrix for a VAR model. We have proposed a method to take
advantage of the trials of an experiment for estimating the parameters
of the VAR model. Our nonparametric estimator is the classical smoothed
periodogram. We have proposed a method to take advantage of the trials
of an experiment to select the optimal smoothing span of the smoothing
kernel. We then outlined a simple method for estimating the optimal
shrinkage weight to construct the generalized shrinkage estimator and
then evaluated its performance on simulated data sets before using it
to analyze functional connectivity in an EEG data set.

The performance of the generalized shrinkage estimator is a function of
the performance of each of the parametric and nonparametric components.
In fact, it can be shown that the risk for the generalized shrinkage
estimator is a weighted average of the risk of each of the two
components less a correction term for the distance between the two
components. If the true spectral density matrix can be well
approximated by that given by a VAR model, then an estimator based on
the VAR model alone will outperform the generalized shrinkage
estimator. However, one can never truly know how well the VAR model
approximates the true spectral density matrix. Our generalized
shrinkage estimator first fits the VAR model, and then adjusts this fit
with the nonparametric smoothed periodogram. The optimal shrinkage
weight is picked by minimizing the risk over all $P$ dimensions of the
multivariate time series simultaneously. This can be problematic if
there is a wide range of dynamics across the dimensions of the
multivariate time series. For instance, if the autospectra for all but
one dimension were flat and there is a sharp peak in that one
dimension, then though the VAR estimator will capture that peak and the
smoothed periodogram oversmooths that peak, the generalized shrinkage
estimator will not give the VAR estimator a lot of weight just to
capture that one peak.

There is still a great amount of work to be done with the generalized
shrinkage estimator. It remains to show the large-sample behavior of
the generalized shrinkage estimator for estimating the spectral density
matrix. Large-sample results were given for the shrinkage estimators
described by \citet{Bohm08} and \citet{Bohm09}. However,
their work
constrained the class of shrinkage targets; \citet{Bohm09} used the
scaled identity matrix as the shrinkage target as a way of regularizing
the smoothed periodogram and the shrinkage target used by \citet
{Bohm08} was specifically the one-factor model. In both works they were
able to show consistency of their shrinkage estimator. In this work,
though we used the VAR model as the shrinkage target, when we developed
the generalized shrinkage estimator we have refrained from imposing
conditions on the shrinkage target, and, in fact, our results on the
optimal shrinkage weight remains valid for any shrinkage target. Recall
that the number of parameters in a $\operatorname{VAR}(K)$ model is of the order
$KP^2$. It may be the case that imposing more constraints to decrease
the parameter space of the VAR model or considering other shrinkage
targets with a low-dimensional parameter space will improve the
performance of the generalized shrinkage estimator. In the future, we
would like to investigate the large-sample performance of the
generalized shrinkage estimator when constraints are imposed on the
shrinkage target.

We do not have asymptotic distributions for the estimated partial
coherence in a frequency band via generalized shrinkage. Test
statistics in the literature have been for partial noncorrelation
between two signals so that the test for zero partial coherence is for
all frequencies. Parametric tests for this null hypothesis have been
provided by, for example, \citet{Dahlhaus97} and \citet{Dahlhaus00}.
Nonparametric tests, on the other hand, are difficult to construct; to
create a boostrap distribution, for instance, one would have to somehow
preserve the correlation structure that exists in the other frequency
bands that are not of interest. One approach is to shuffle the data
across trials in order to completely destroy the correlation structure,
but tests on partial coherence over a frequency band using this
approach will have a larger Type I error than advertised. However, one
can take advantage of the multiple trials in the experiment and the
multiple experimental conditions to investigate the differences in
connectivity across the experimental conditions using nonparametric
tests, as we have done here using the jackknife.

\begin{appendix}

\section*{Appendix: Simulation settings}\label{app}

The coefficient matrix for the VMA process is as follows. First, let
\[
\mathbf{\theta}_1 =
\pmatrix{
0 & 0.20 & 0.15 & 0.15 & 0 & -0.15\cr
0.20 & 0 & -0.20 & 0 & 0 & 0\cr
-0.15 & 0.20 & 0 & 0 & 0 & 0 \cr
0 & 0 & 0 & 0 & 0.20 & 0.15 \cr
0 & 0 & 0 & 0.20 & 0 & -0.20 \cr
0 & 0 & 0 & -0.15 & 0.20 & 0
}.
\]
Then the coefficient matrix is
\begin{eqnarray*}
\mathbf{\Theta}_1 =
\pmatrix{
\theta_1 & \mathbf{0}_6 \cr
\mathbf{0}_6 & \theta_1
}
.
\end{eqnarray*}

The coefficient matrices for the VAR process are as follows. Let
\[
\phi_1 = (0.75, 0.75, 0.75).
\]
Then the two coefficient matrices are
\begin{eqnarray*}
\mathbf{\Phi}_1 &=& \operatorname{diag}(\phi_1, \phi_1, \phi_1, \phi_1),
\qquad
\mathbf{\Phi}_2 = -0.20 \cdot\mathbf{I}_{12},\qquad
\mathbf{\Phi}_3 = \mathbf{0}_{12},
\\
\mathbf{\Phi}_4 &=& -0.15 \cdot\mathbf{I}_{12} \quad\mbox{and} \quad
\mathbf{\Phi}_5 = -0.05 \cdot\mathbf{I}_{12}.
\end{eqnarray*}
The noise process $\mathbf{Z}(t)$ is a zero-mean 12-dimensional
Gaussian process with variance--covariance matrix $\mathbf{I}_{12}$. A
realization $\mathbf{X}(t)$ of the mixture process takes the form
$\mathbf{X}(t) = 0.65 \cdot\mathbf{X}_{\mathrm{MA}}(t) + 0.35 \cdot
\mathbf{X}_{\mathrm{AR}}(t)$, where $\mathbf{X}_{\mathrm{MA}}(t)$ is a~%
VMA and $\mathbf{X}_{\mathrm{AR}}(t)$ is a VAR, and the two are
independent. Because the two processes are independent, then the
spectral density matrix of the mixture process is the weighted sum of
the spectral density matrix of the VMA process and the spectral density
matrix of the VAR process.

\end{appendix}

\section*{Acknowledgments}
%This research is supported in part by the National Science Foundation
%(Division of Mathematical Sciences).
The authors would like to thank
Jerome N. Sanes (Neuroscience, Brown University) for sharing the EEG
data set. The authors would also like to thank the Editor and the
anonymous Associate Editor and referee for their suggestions that have
led to an improved paper.

%%%%%%%%%%%%%%%%%%%%%%%%%%%%%%%%%%%%%%%%%%%%%%%%%%%%%%%%%%%%%%%%%%%%%%%%%%%%%%%%%%%%%%%%

%suskaldyti doi

\printaddresses

\end{document}